\documentclass[aps,pra,preprint,superscriptaddress,floatfix]{revtex4}

\usepackage{amsmath}
\usepackage{graphicx}
\usepackage{dcolumn}
\usepackage{subfigure}
\usepackage{mathrsfs}
\usepackage{multirow}
\usepackage{mciteplus}
\usepackage{color}
\usepackage{ulem}

\begin{document}


\title{Large-scale atomistic density functional theory calculations of phosphorus-doped silicon quantum bits}


\author{Loren Greenman}
\affiliation{Department of Chemistry and Kenneth S. Pitzer Center for Theoretical Chemistry, University of California, Berkeley, CA 94720, USA}
\author{Heather D. Whitley}
\affiliation{Condensed Matter and Materials Division, Lawrence Livermore National Laboratory}
\altaffiliation[This author's current location is\ ]{AX Division, Lawrence Livermore National Laboratory}
\author{K. Birgitta Whaley}
\affiliation{Department of Chemistry and Kenneth S. Pitzer Center for Theoretical Chemistry, University of California, Berkeley, CA 94720, USA}
\email{whaley@berkeley.edu}


\date{\today}


\pacs{}

\begin{abstract}
We present density functional theory calculations of phosphorus dopants in bulk silicon and of several properties relating to their use as spin qubits for quantum computation.
\textcolor{black}{Rather than a mixed pseudopotential or a Heitler-London approach,} we have used an explicit treatment for the phosphorus donor and 
examined the detailed electronic structure of the system as a function of the isotropic doping fraction, including lattice relaxation due to the 
presence of the impurity.
Doping electron densities ($\rho _{doped} - \rho _{bulk}$) and spin densities ($\rho _{\uparrow} - \rho _{\downarrow}$) are examined in order to study the properties of the 
dopant electron as a function of the isotropic doping fraction.
Doping potentials ($V_{doped} - V_{bulk}$) are also calculated for use in 
calculations of the scattering cross-sections of the phosphorus dopants, which are important in the 
understanding of electrically detected magnetic resonance experiments.
We find that the electron density around the dopant leads to non-spherical features in the doping potentials, such as trigonal lobes in the (001) plane at 
energy scales of +12 eV near the nucleus and 
of -700 meV extending away from the dopants.  
These features are generally neglected in effective mass theory 
and will affect the coupling between the donor electron and the phosphorus nucleus.  
Our density functional calculations reveal detail in the densities and potentials of the dopants which are not evident in calculations that do not include explicit treatment of the 
phosphorus donor atom and relaxation of the crystal lattice.
These details can also be used to parameterize tight-binding models 
for simulation of large-scale devices.
\end{abstract}

\maketitle


\section{Introduction}

Dopants in silicon show potential as qubits for solid-state quantum computers~\cite{Kane_Proposal,Skinner_Proposal,Loss_DiVincenzo_Proposal}, with the advantages of scalability as well as the promise of utilizing the 
existing semiconductor industry and its processing techniques~\cite{Kane_Proposal,Skinner_Proposal,Loss_DiVincenzo_Proposal,Morton_Nature_Review,Levy_JPCM_Review}.
The theory of Group V dopants 
such as phosphorus in silicon~\cite{KohnLuttinger} is useful for describing the quantum nature of the electrons in these systems, as well as for developing schemes to circumvent one of the most challenging aspects of solid-state quantum computers, namely environmental decoherence~\cite{Morton_Nature_Review,Feher_ESR1,Feher_ESR2,Gordon_PRL_58,Witzel_PRB_06,Abe_PRB_10,deSousa_PRB_03,Tyryshkin_PRB_03,Schenkel_APL_06}.
In order to provide a benchmark for such theories and also to use as a starting point for building efficient and accurate tight binding methods, an \textit{ab initio} description of 
dopants in silicon is desired.
However, the size of the systems required to describe doped silicon at or near the single-dopant limit is large, making such a description computationally expensive.
In this work, we present large-scale density functional theory calculations for phosphorus-doped silicon supercells with up to 432 atoms.
We make comparisons to other theoretical works~\cite{Sarma_Heitler_London,Carter_Nanotech_11,Kettle_JPCM_04,Wellard_PRB_03}  to determine what can and cannot be captured by approximate or single-electron theories for the doped-silicon systems.

Previous efforts to describe the electronic structure of silicon dopants include effective mass approaches beginning with the work of Kohn and Luttinger~\cite{KohnLuttinger} and continuing with many others~\cite{MacMillen_PRB_84,Kettle_PRB_03,Kettle_JPCM_04,Wellard_JPCM_04,Calderon_PRL_06,Calderon_PRB_07,Hao_PRB_09}, including Fang \textit{et. al.} who perform two-electron Hartree-Fock calculations within effective mass theory~\cite{Fang_PRB_02}. 
These efforts include calculating the effects of applied electric and magnetic fields~\cite{Kettle_PRB_03,Calderon_PRL_06} and the coupling of two donors via the exchange interaction~\cite{Kettle_JPCM_04,Wellard_PRB_03,Wellard_JPCM_04,Calderon_PRB_07}.
Tight-binding calculations have also been performed~\cite{Lansbergen_NP_08,Rahman_PRB_09,Rahman_PRB_10}, including a calculation of the quadratic Stark coefficient of the hyperfine interaction which 
has reproduced experimentally measured values more accurately than effective mass
theory~\cite{Rahman_PRL_07}.
\textcolor{black}{Two-dimensional layers of dopants in silicon known as $\delta$-layers have been described using density functional theory with compact atomic-orbital basis sets~\cite{Carter_PRB_09,Carter_PRB_09E}, and additional DFT studies~\cite{Carter_Nanotech_11,Carter_PRB_13} evaluated the use of mixed pseudopotentials, which treat the dopant and silicon atoms in the layer using the same core potential, and compared them to all-atom calculations.}
These DFT calculations and a number of additional calculations (see \textcolor{black}{Ref.~\cite{Drumm_NRL_13} and Refs. [30-40] therein}) show a large amount of disagreement for calculated properties such as the valley splitting.
\textcolor{black}{Although some of this disagreement can be attributed to geometrical effects of dopant placement which we will not explore in detail (see instead Refs.~\cite{Drumm_NRL_13}~and~\cite{Carter_Nanotech_11}), the accuracy of the description of dopant electronic structure contributes to these discrepancies.}
Additionally, electrically detected magnetic resonance (EDMR)~\cite{Lo_PRL_11} has called into question a theoretical picture of the scattering of electrons in a two-dimensional electron gas (2DEG) from dopants in silicon~\cite{Ghosh_PRB_92,deSousa_PRB_09}.
An \textit{ab initio} description of a dopant in silicon is therefore useful both as a benchmark and for determining the details of the electronic structure of an isolated dopant which can subsequently be used to calculate more accurate spin-dependent scattering cross sections.
Although these are expensive calculations, we have been able to perform large-scale calculations using the computational resources at Lawrence Livermore National Laboratory.

\section{Methods}
We have used the {\sc Quantum ESPRESSO} suite of programs~\cite{QE-2009} in order to perform density functional theory calculations using a basis set of plane waves.
Face-centered cubic lattices of substitutionally doped silicon were prepared at variable dopant ratios by substituting 1 phosphorus atom in unit cells with 53, 127, 249, and 431 silicon atoms, respectively.  We used the 
Perdew-Burke-Ernzerhof (PBE) density functional~\cite{PBE1,PBE2} with an ultrasoft pseudopotential for phosphorus (P.pbe-n-van.UPF from Ref.~\cite{qewebsite}) and a norm-conserving 
pseudopotential that was calculated using FHI98PP\cite{Fuchs99} for silicon.  
A plane-wave energy cutoff of 65~Ry. was chosen based on convergence of the total energy and pressure in the smallest supercell. 
K-space sampling was performed using a Monkhorst-Pack~\cite{MonkhorstPack} grid of 8$\times$8$\times$8 (54 atom), 6$\times$6$\times$6 (128 atom), 4$\times$4$\times$4 
(250 atom), and 2$\times$2$\times$2 (432 atom) grid points.
As a reference for the energy and properties of bulk silicon, a two-atom silicon cell was used with a grid of 20$\times$20$\times$20 k-points.
In this cell, one silicon is placed at the origin and another is placed at the point $(\frac{a}{4},\frac{a}{4},\frac{a}{4})$, where $a$ is the lattice constant.
The basis vectors of the FCC cell are $(\frac{a}{2},\frac{a}{2},0)$, $(\frac{a}{2},0,\frac{a}{2})$, and $(0,\frac{a}{2},\frac{a}{2})$; the two silicon atoms are repeated at every integer multiple of the basis vectors.
The lattice constant of the doped supercells was determined as multiples of the 5.46 \AA~lattice constant for bulk silicon computed with the pseudopotential used in this study.  
\textcolor{black}{
The phosphorus dopants repeat along the directions of the basis vectors.  
The directions of the plots given below were chosen to be orthogonal to each other.  
Therefore, for the (001) plane, first the [100] and [010] basis vectors were chosen, and then the Gram-Schmidt procedure was used to find the orthogonal direction ([-1 2 0]).  
The angular bracket notation is used in the plots because the choice of direction is somewhat arbitrary due to the FCC symmetry of the system.
}
Geometric minimization of the total energy was performed for all doped systems studied here. In the calculations with $N<432$ atoms, all of the atoms were 
allowed to move during the simulation, while for $N=432$, the atoms at the edges of the supercell were frozen in order to make better comparisons to the bulk silicon system.  The 
volume of each supercell was constant during the geometric optimization.   
These computations required 600 processors on the Ansel supercomputer at Lawrence Livermore National Laboratory, a 324 node system using the Intel Xeon architecture rated at a peak of 43.5 TFLOP/s~\cite{llnlcomputing}.
A single point energy calculation for the 432-atom cell took about 19 CPU hours to complete.

In Section~\ref{potentialsection}, we use the doping potentials as a measure of the electronic environment of the dopants.
The electric potential for each cell is obtained by adding the nuclear term ($V_{nuc}$), coulombic Hartree term ($V_{Hartree}$), and PBE exchange-correlation term ($V_{xc}$) at the converged electronic density $\rho _{conv}$,
\begin{equation}
\label{cellpotential}
V = V_{nuc} + V_{Hartree} [\rho _{conv}] + V_{xc} [\rho _{conv}].
\end{equation}
The density $\rho _{conv}$ is that which minimizes the energy of the system according to Eq.~(\ref{cellpotential}).
The doping potential is then obtained by subtracting the potential obtained for the undoped cell from that of the doped cell,
\begin{equation}
\label{dopingpot}
V_{doping} = V_{doped} - V_{undoped}.
\end{equation}
Similarly, the doping density discussed in Section~\ref{densitysection} is obtained by subtracting $\rho _{conv}$ of the undoped cell from that of the doped cell.
Finally, the spin densities discussed in Section~\ref{spinsection} are obtained by subtracting the spin-up and spin-down portions of the density $\rho _{conv}$ of a single cell.

In order to estimate the exchange coupling (J), we use the DFT broken spin symmetry states~\cite{Rudra_JCP_06,Ruiz_JCC_11,Yamaguchi_CPL_79,Noodleman_JACS_85,Noodleman_JCP_81},
\begin{equation}
\label{jcoupling}
\frac{J}{2}=E_{BS}-E_{HS},
\end{equation}
where $BS$ and $HS$ denote, respectively, the broken symmetry ground state and high-spin state in which the spin density is constrained to give a spin of $S=\frac{1}{2}$ in each unit cell.
This procedure empirically has given exchange couplings with a satisfactory degree of accuracy for molecular systems~\cite{Rudra_JCP_06,Ruiz_JCC_11}, and the exchange couplings it calculates can be thought of as the spin-coupling parameter $J$ in both Ising and Heisenberg models of the spin interactions in the periodic system~\cite{Illas_TCA_00,Dai_JCP_03}.

\section{Results and Discussion}
The Bohr radius of phosphorus in silicon is estimated to be about 2.5 nm~\cite{KohnLuttinger}.
In order to reach the single dopant limit, the phosphorus atoms in a silicon matrix would have to be spaced a minimum of $\sim5$~nm apart.
This would require a supercell of tens of thousands of atoms and is outside the realm of feasibility for most DFT calculations.  
Here, we have therefore assessed the energies and properties of doped silicon as a function of the doping fraction approaching this limit, in order to develop a better 
understanding of how explicitly treating the phosphorus donor atom and lattice relaxation effects the electronic structure.  
\textcolor{black}{In Table~\ref{energytable}, we 
show the convergence (as a function of the size of the supercell) of 
the geometrically relaxed and unrelaxed formation energy,
\begin{equation}
\label{formenergy}
E_{doped}-(E_P+(N_{atom}-1)E_{Si}) - (E_{undoped}+N_{atom}E_{Si}),
\end{equation}
where $E_{doped}$ is the energy of the doped supercell at either the geometrically optimized (relaxed) or bulk silicon (unrelaxed) geometry, $E_P$ is the energy of an isolated phosphorus atom, $E_{Si}$ is the energy per atom of bulk silicon, $E_{undoped}$ is the energy of the undoped supercell of $N_{atom}$ atoms, and the last term in parentheses in Eq.~(\ref{formenergy}) is included to account for using approximate energy cutoffs and k-point grids for the larger supercells.
\begin{table}[htp!]
\color{black}
\caption{Convergence of the energy as a function of unit cell size.
For each unit cell from 54 to 432 atoms, we give the unrelaxed formation energy (column 2), the relaxation energy (column 3), and the relaxed formation energy (column 4)
\label{energytable}}
\begin{ruledtabular}
\begin{tabular}{cddd}
\multicolumn{1}{c}{Unit cell}&	
\multicolumn{1}{c}{Formation Energy (unrelaxed)}&
\multicolumn{1}{c}{Relaxation Energy}&
\multicolumn{1}{c}{Formation Energy (relaxed)}\\
&
\multicolumn{1}{c}{(eV)}&
\multicolumn{1}{c}{(meV)}&
\multicolumn{1}{c}{(eV)}\\
\hline
Si$_{53}$P&-3.1519&-30.0512&-3.1820\\
Si$_{127}$P&-3.1816&-27.8154&-3.2095\\
Si$_{249}$P&-3.1872&-32.1070&-3.2193\\
Si$_{431}$P&-3.2322&-32.3320&-3.2646
\end{tabular}
\end{ruledtabular}
\end{table}
The geometric relaxation energy of the doped system (column 3 of Table~\ref{energytable}) is similar (around 30~meV) for each of the cells.
The formation energies increase in magnitude as a function of cell size, suggesting that the presence of the defect is reducing the strain in the lattice.
The relaxed formation energy increases in magnitude by 82.6 meV from the 54 atom cell to the 432 cell and by 45.3 meV from the 250 to the 432 atom cell.  
This suggests that the effects of lattice relaxation and changes in the electronic structure will be important for the donor electron dynamics in phosphorus doped silicon where 
the isotropic doping fraction is $\sim 0.2$~\% or higher.}  In systems with a lower doping fraction, these effects may still play a role, but we cannot make a definitive statement on this issue 
since the changes in energy exhibit a strong dependence on the system size up to the largest system studied here.  In another recent study\textcolor{black}{~\cite{Drumm_NRL_13}}, the energy gained by relaxing a cell with a monolayer of phosphorus 
dopants was found to be of a similar magnitude.  Whether the lattice relaxation is important to EDMR readout schemes depends on how the relaxation affects the scattering dynamics of electrons, which is related to the doping potential.
In the next section, we make comparisons of the doping potentials 
for a donor phosphorus atom in silicon, both with and without the effects of geometrical relaxation of the lattice. 

\subsection{Doping potential\label{potentialsection}}
We define the doping potential as the doped cell potential minus the bulk silicon potential (Eqs.~(\ref{cellpotential})~and~(\ref{dopingpot})).
The doping potential shows how the electronic environment surrounding the dopant differs from that of bulk silicon.
These potentials also provide input for calculations of the cross sections of electron scattering at the dopants~\cite{Lordi_Scattering_PRB_10,SCARLET},
which are largely determined by integrals of the doping density~\cite{Lordi_Scattering_PRB_10}.
By calculating the scattering of conduction electrons confined in a two-dimensional layer located at a given distance from the (001) plane, a connection can be made with electrically detected magnetic resonance (EDMR) schemes~\cite{Ghosh_PRB_92,Xiao_Nature_04,Lo_PRL_11,vanBeveren_APL_08,Lo_APL_07} used to measure the dopant spin state.
Doping potentials calculated using atomistic DFT can also be used to parameterize new tight binding models, or effective single-electron models which more accurately reproduce the effects of the dopant electronic structure than standard effective mass models.

\begin{figure}[htp!]
\subfigure[]{\includegraphics[height=2in,width=2in,angle=0.0]{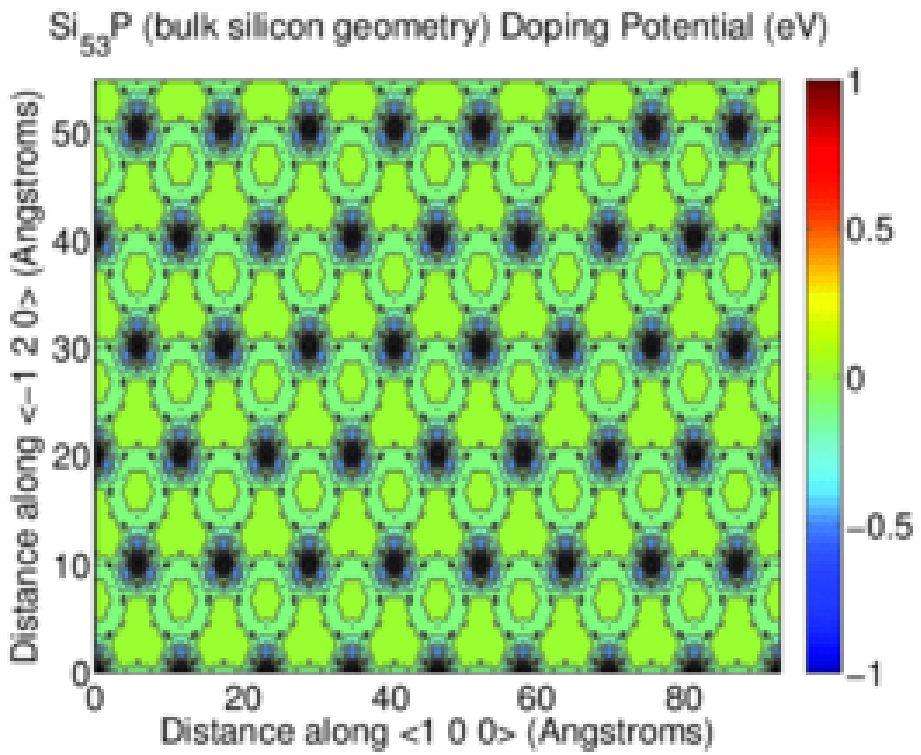}\label{m54u}}
\subfigure[]{\includegraphics[height=2in,width=2in,angle=0.0]{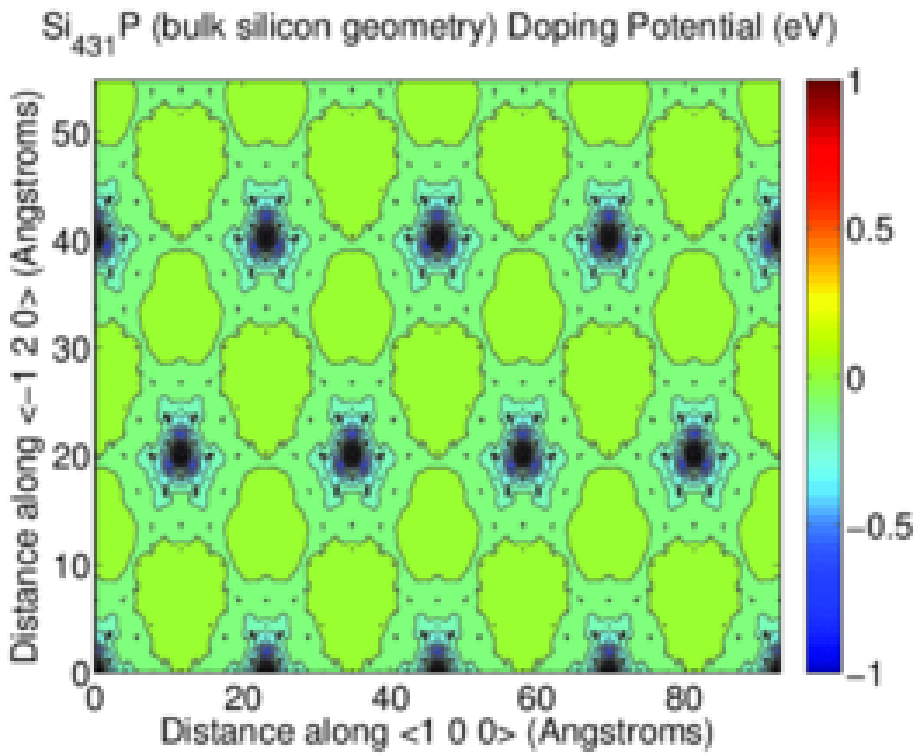}\label{m432u}}\\
\subfigure[]{\includegraphics[height=2in,width=2in,angle=0.0]{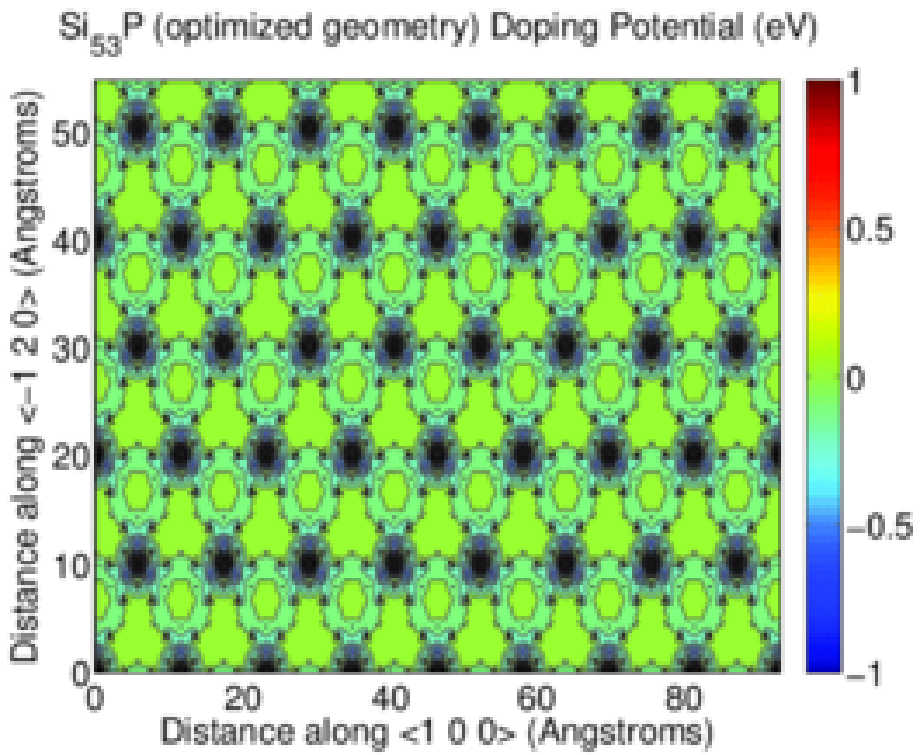}\label{m54}}
\subfigure[]{\includegraphics[height=2in,width=2in,angle=0.0]{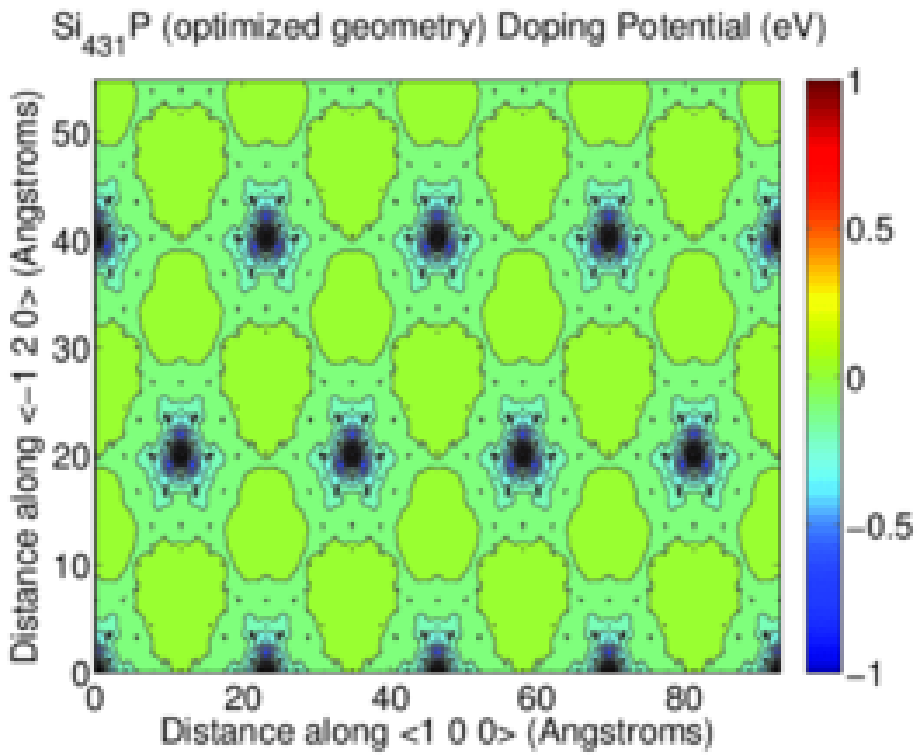}\label{m432}}
\caption{(Color online) The potential difference (eV) in the (001) plane between doped and undoped silicon is shown for cell sizes of 54~(\ref{m54u}~and~\ref{m54}) and 432 atoms~(\ref{m432u}~and~\ref{m432}).
An area of 93 $\times$ 55 \AA  is shown for in all figures, in which the 54 and 432 atom cells repeat about 44 and 11 times, respectively.
For the 54 and 432 atom cells, parts of 51 and 14 dopants, respectively, can be seen in the figures.
Doping potentials are shown for cells at the bulk geometry~(\ref{m54u}~and~\ref{m432u}) and optimized geometries~(\ref{m54}~and~\ref{m432}).
The contour lines are drawn every 1 eV except between -2 and 2 eV, where they are drawn every 100 meV.
The color axis in these figures is between -1 and 1 eV in order to highlight the effects at this energy scale, while the range of doping energies is between -9 and 12 eV, with the larger values occuring near the dopant nucleus (see Fig.~\ref{potentialcufigure}).
\label{potentialfigure}}
\end{figure}

\begin{figure}[htp!]
\subfigure[]{\includegraphics[height=2in,width=2in,angle=0.0]{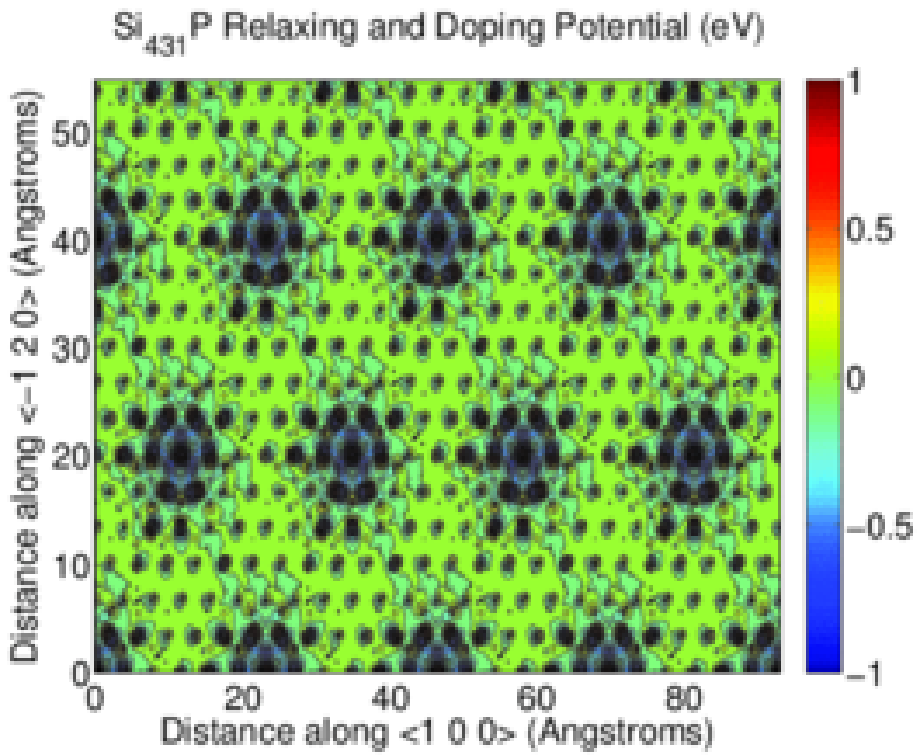}\label{m432rd}}
\subfigure[]{\includegraphics[height=2in,width=2in,angle=0.0]{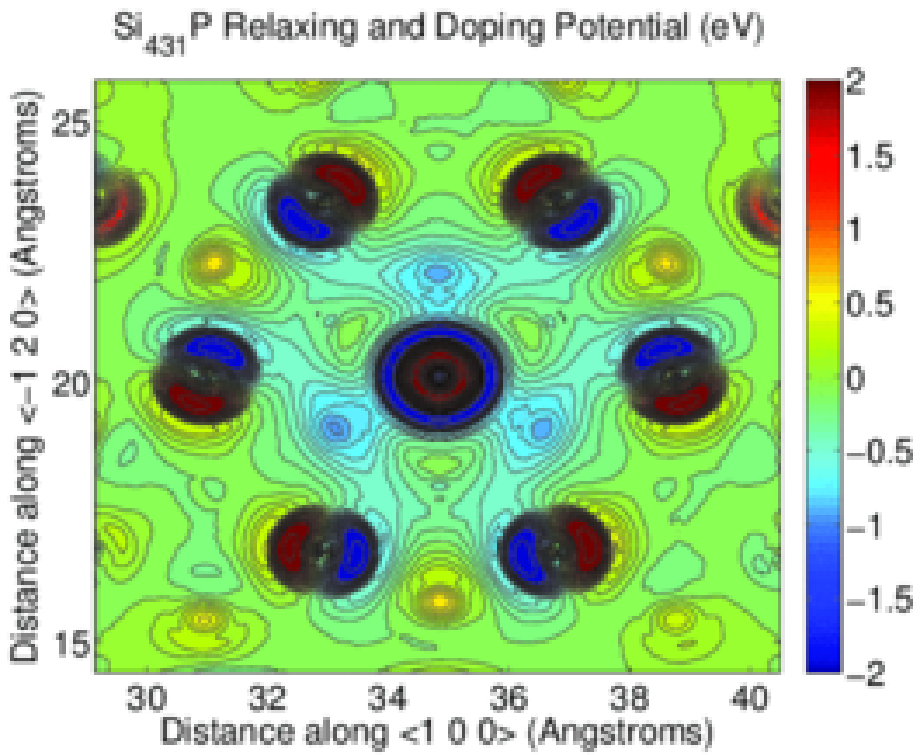}\label{m432rdz}}
\caption{(Color online) The potential difference (eV) in the (001) plane between doped silicon in its optimized geometry and bulk silicon for the 432 atom cell in the bulk geometry.
We refer to this as the ``relaxing and doping potential.''
In Fig.~\ref{m432rd}, 14 phosphorus dopants are visible, while Fig.~\ref{m432rdz} shows a close-up of the region around one dopant.
The contours and color axis in Fig.~\ref{m432rd} are the same as in Fig.~\ref{potentialfigure}.
The close up in Fig.~\ref{m432rdz} uses the same contours, but the color axis is between -2 and 2 eV.
\label{relaxedpotfigure}}
\end{figure}

\begin{figure}[htp!]
\subfigure[]{\includegraphics[height=3in,width=3in,angle=0.0]{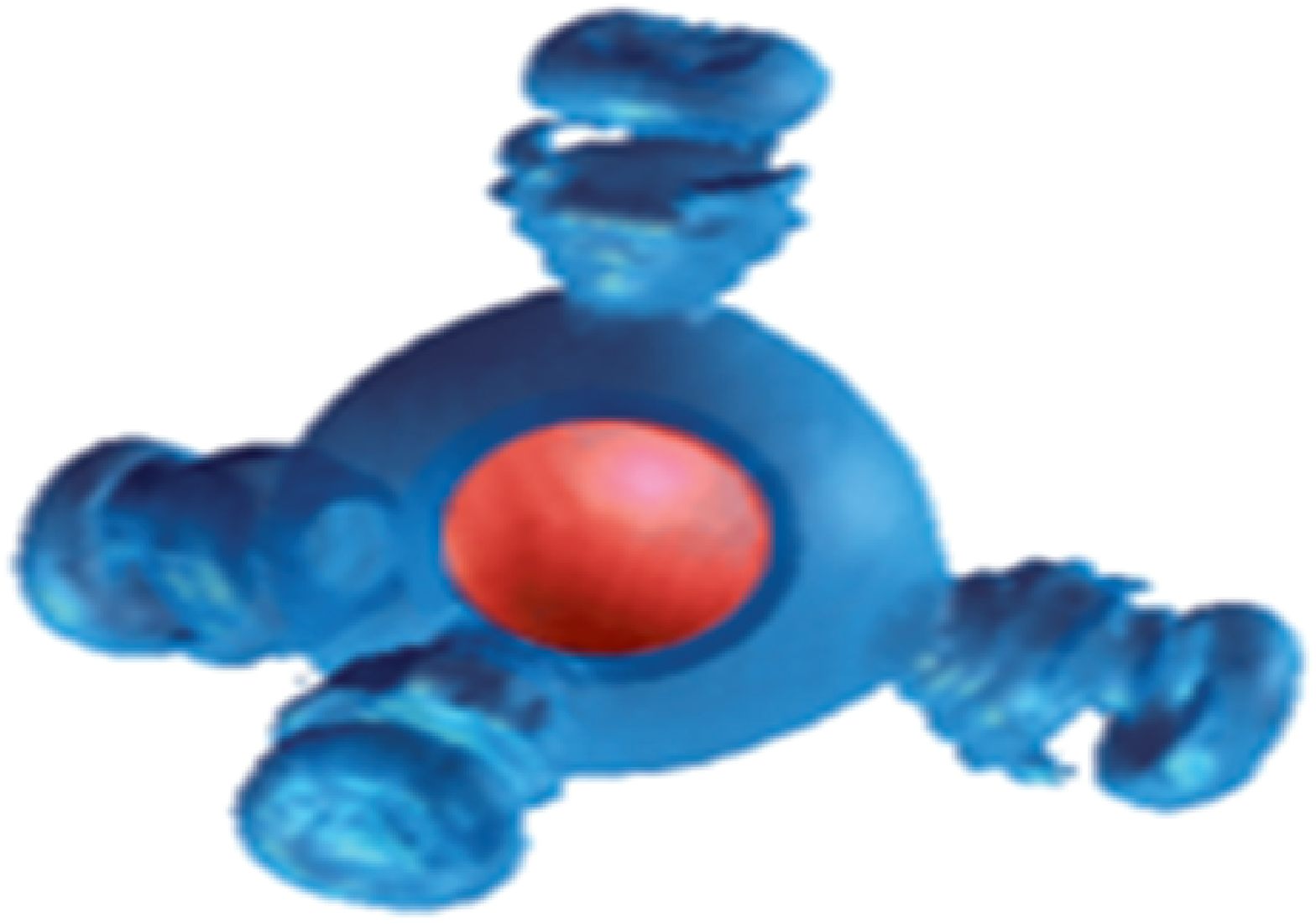}\label{3d700}}
\subfigure[]{\includegraphics[height=3in,width=3in,angle=0.0]{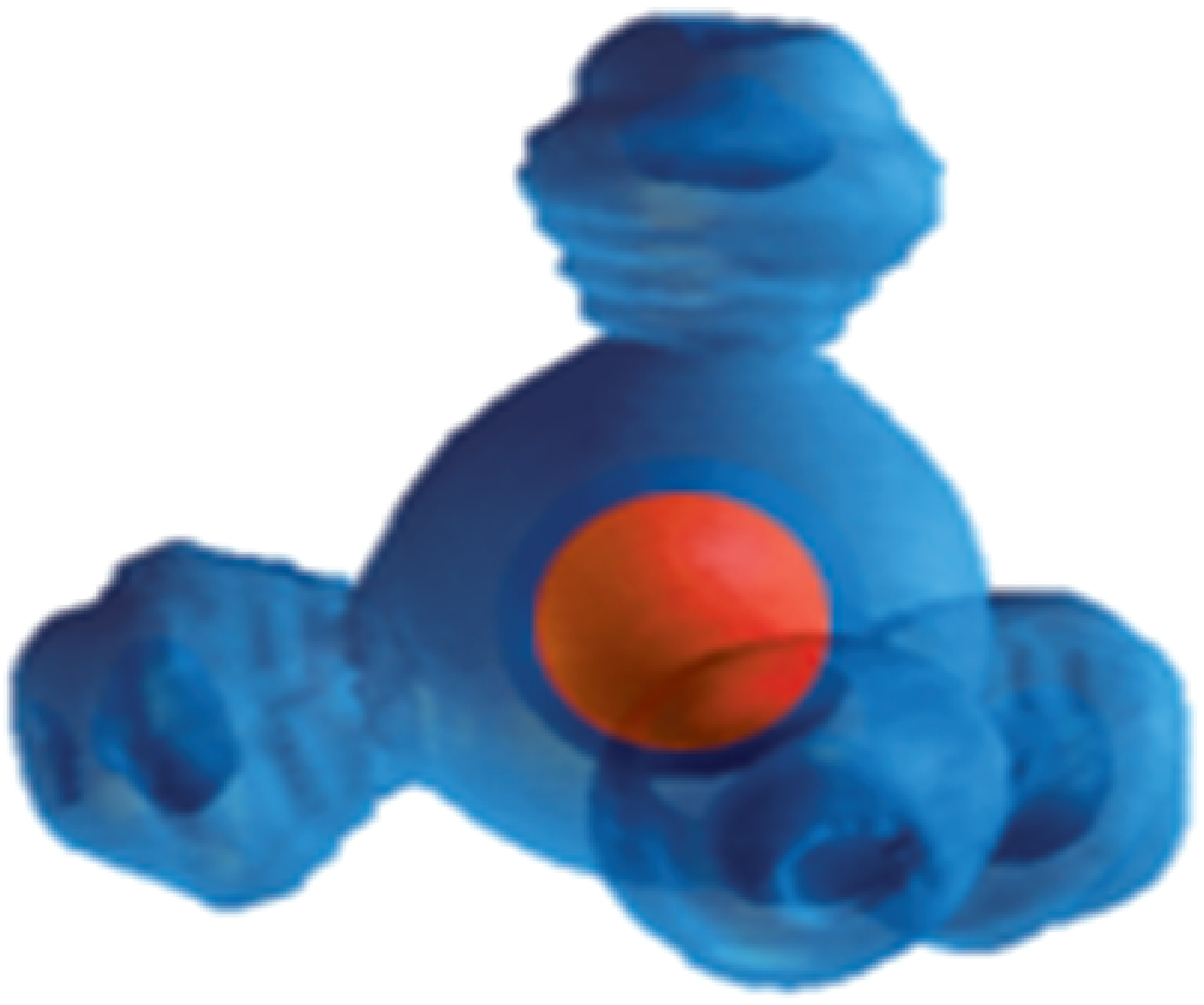}\label{3d600}}\\
\subfigure[]{\includegraphics[height=3in,width=3in,angle=0.0]{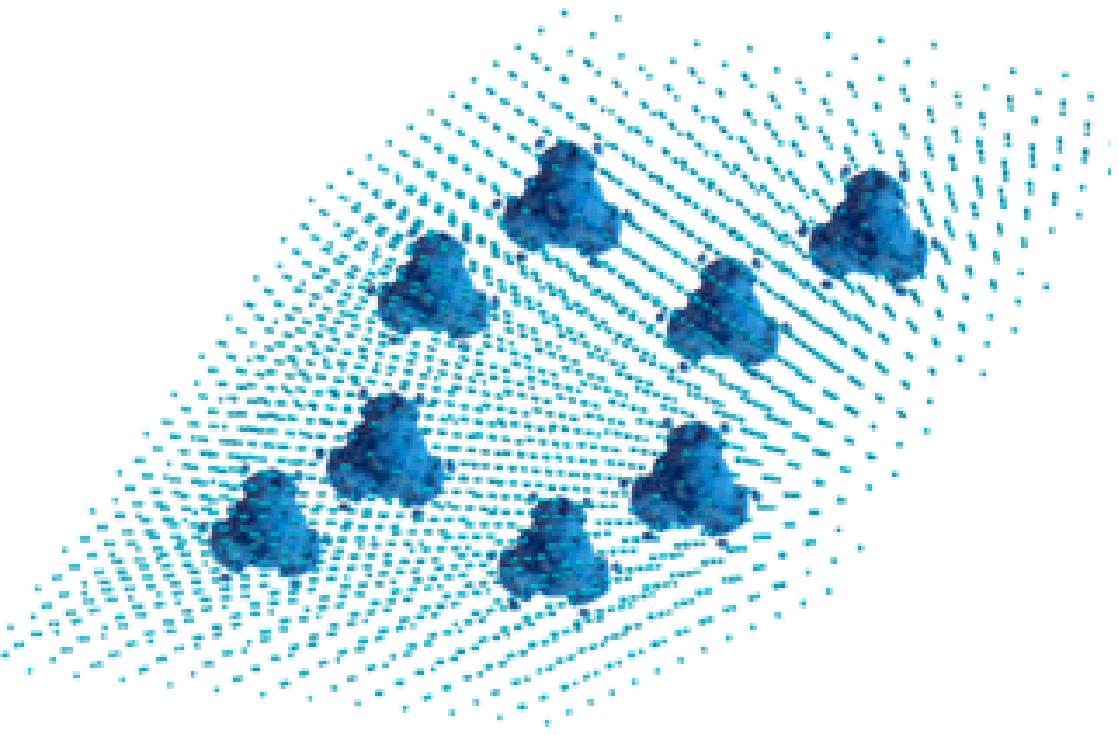}\label{3d150}}
\caption{(Color online) Doping potential isosurfaces are shown for -700~(\ref{3d700}), -600~(\ref{3d600}), and -150~(\ref{3d150}) meV.
The isosurfaces are shown in slightly transparent blue, and the phosphorus donors are shown in red.
The silicon atoms are not pictured in Figs.~\ref{3d700}~and~\ref{3d600}, and they are shown for reference in Fig.~\ref{3d150}.
\label{3dpotentialfigure}}
\end{figure}

The doping potentials  are shown for the 54 and 432 atom cells in Figs.~\ref{potentialfigure}~and~\ref{relaxedpotfigure}.
\textcolor{black}{In Fig.~\ref{potentialfigure}, the potentials are given for both the doped and undoped cells at geometrically unrelaxed (Figs.~\ref{m54u}~and~\ref{m432u}) and relaxed (Figs.~\ref{m54}~and~\ref{m432}) geometries.  In Fig.~\ref{relaxedpotfigure}, the effects of geometric relaxation on the doping potentials is given.}
In the (001) plane, the doping potentials for the 54 atom cell (left panels) can be seen to overlap.  
Thus, while there are areas in which the doping potential goes to zero, indicating a return to bulk silicon behavior, the dopants are largely connected by regions of potential greater than 100 meV.
In the 432 atom cell at the bulk geometry, the doping potentials nearly go to zero between the dopants. 
However, there are small areas of non-zero potential still connecting dopants, suggesting that the single dopant limit has still not been reached.
These potential connections between dopants are slightly exaggerated at the relaxed geometry.
In both cells, however, the potential region directly around the dopants is similar, with a region exceeding 8 eV directly around the dopant and three 1 eV lobes about 120 degrees apart from each other.
\textcolor{black}{These lobes are evident even as geometric relaxation effects are included (Fig.~\ref{relaxedpotfigure}).  Additional oscillations are evident in the potential when geometric relaxation effects are included as a result of the position of the nearby silicon nuclei, although the pattern of oscillations near the dopant are similar with and without relaxation effects.}
In addition to the lobes near the nucleus, there are additional lobes at an energy scale of -700 meV which extend away from the nuclei in space.
The trigonal symmetry results from taking a two-dimensional cut through the three dimensional structure onto the (001) plane.
When viewed in three dimensions, these lobes can be seen to arise from the tetrahedral nature of the bonding in the silicon cell.
Three-dimensional potential isosurfaces at -700, -600, and -150 meV are shown in Fig.~\ref{3dpotentialfigure}.
\textcolor{black}{
To address the question of whether the tetrahedral doping potential configuration arises from the FCC structure of the periodic cell images, we have also performed calculations for rectangular cells with an aspect ratio of two to one.
The dopant potentials for these cells (not pictured) exhibit tetrahedral doping potentials similar to those seen in Figs.~\ref{potentialfigure},~\ref{relaxedpotfigure},~and~\ref{3dpotentialfigure}.
This suggests that the salient features of the doping potential such as the tetragonal lobes are not dependent on the geometry of the cell arrays.
}
The non-spherical nature of the doping potential will be important for calculation of electron-dopant scattering cross sections.
Effective mass models of doped silicon assume a spherical Coulomb potential, while some tight binding models use a spherical, Coulomb-like doping potential~\cite{NEMO3D,Rahman_PRL_07,Rahman_PRB_10,Rahman_PRB_11} and allow the surrounding atoms' electronic structure to adjust according to this potential.
The anisotropic doping potentials and cell geometries calculated here could  be used as alternative parameterizations for tight binding models, and they can also be used to calibrate the resulting potentials and densities calculated by tight binding models.

\begin{figure}[htp!]
\subfigure[]{\includegraphics[height=3in,width=3in,angle=0.0]{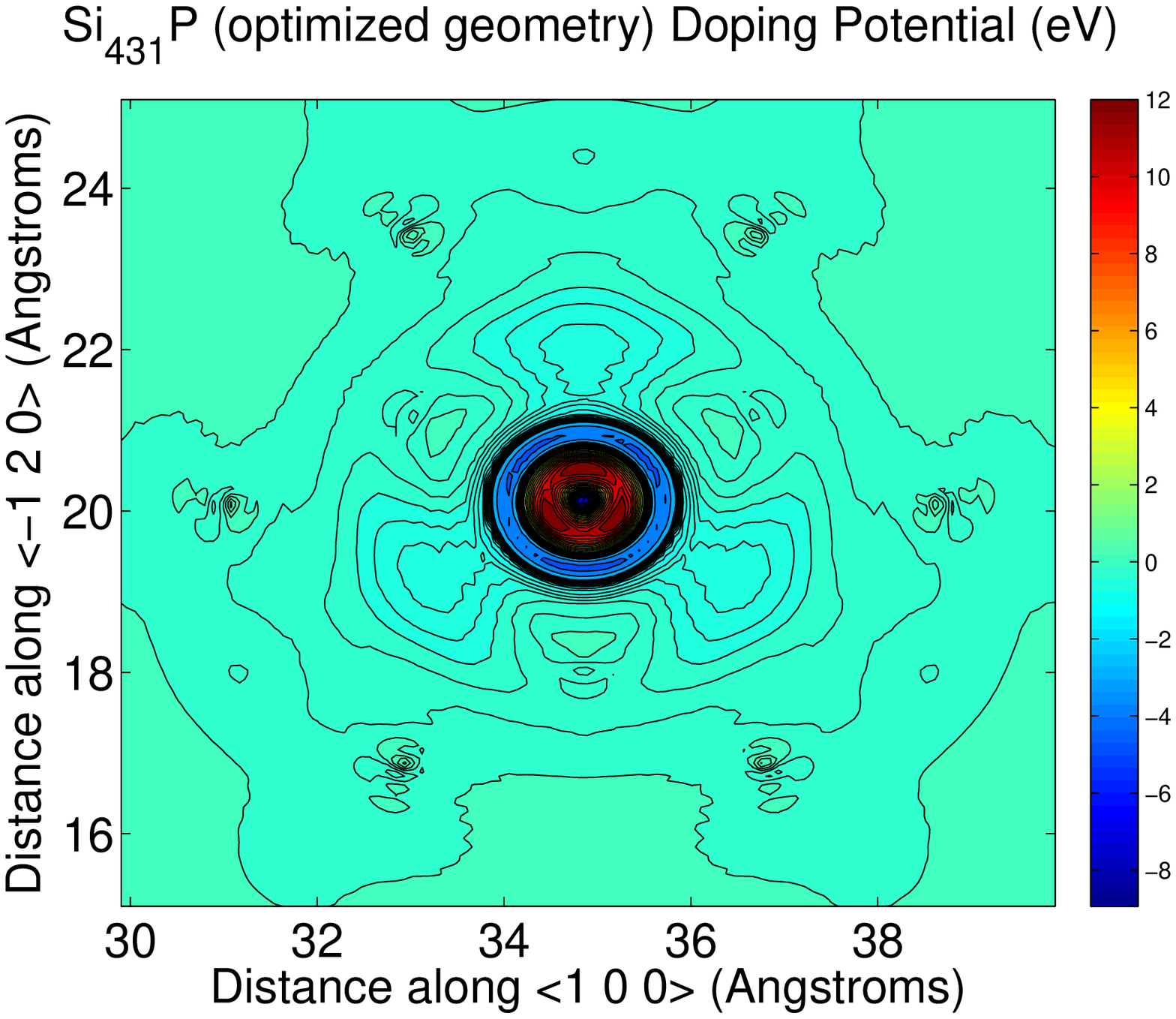}}
\subfigure[]{\includegraphics[height=3in,width=3in,angle=0.0]{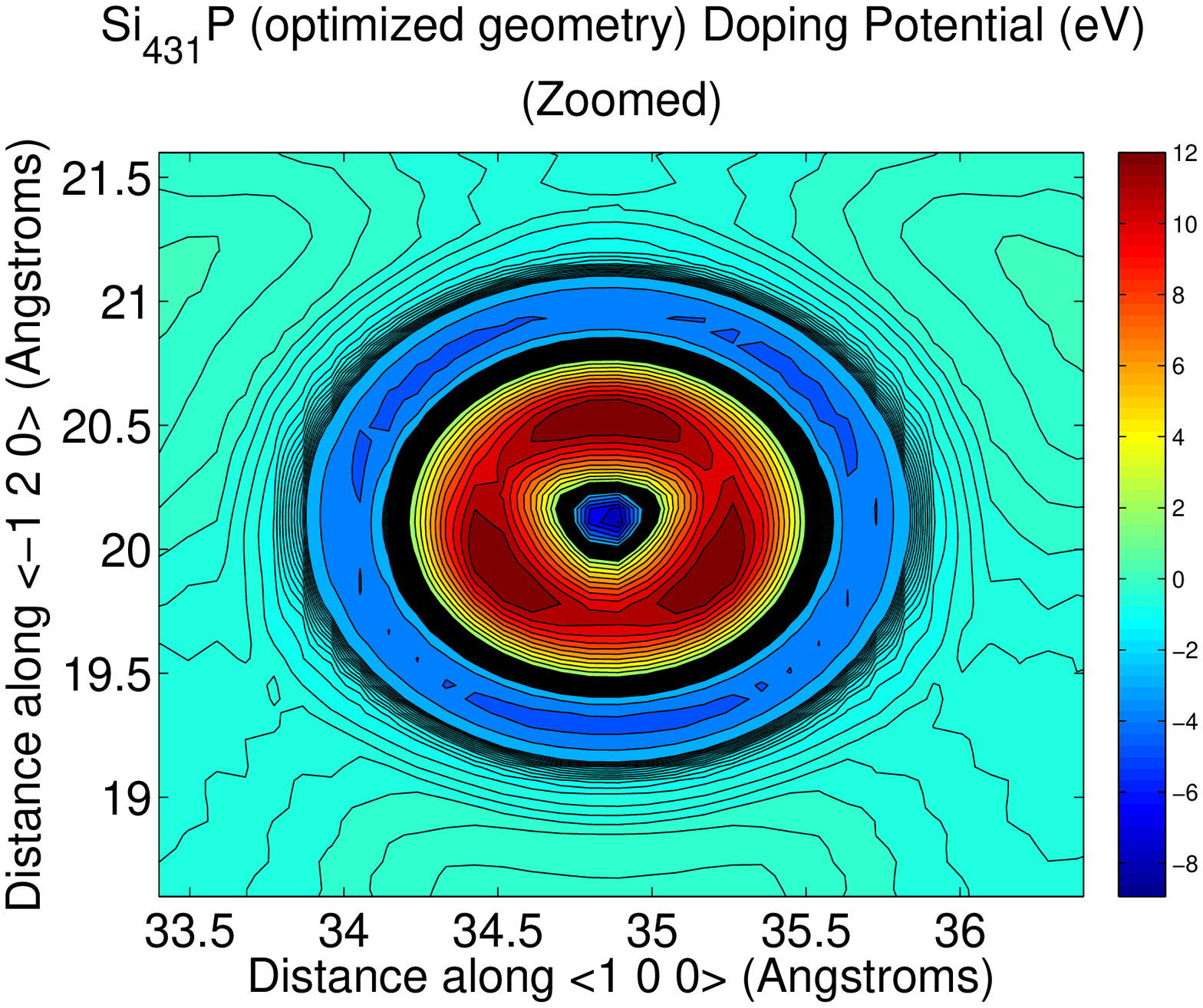}}
\caption{(Color online) The potential difference (eV) between doped and undoped silicon is shown for the 432 atom cell at the optimized geometry of the doped cell for the region close to a  dopant.
\textcolor{black}{Different zooms are shown in panels (a) and (b).}
The doping potential can be seen to oscillate in this region as opposed to showing s-orbital character as predicted by effective mass theory.
Additionally, there are a number of features which are not spherically symmetric.
The contour lines are the same as in Fig.~\ref{potentialfigure}, while the color axis has been expanded to the range of -9 to 12 eV.\label{potentialcufigure}}
\end{figure}

Fig.~\ref{potentialcufigure} shows a closer view of the doping potential in the region of the dopant for the 432 atom cell with optimized geometry.
In the effective mass picture~\cite{KohnLuttinger}, the dopant wavefunction has s-orbital character.
In Fig.~\ref{potentialcufigure}, the potential in the region of the dopant can be seen to oscillate as a function of the orientation.
The oscillations are due to interactions with electrons in the shells below the valence shell.
If these calculations were performed without using pseudopotentials, which reduce oscillations from core electrons and replace them with a smooth potential, the potential would most likely oscillate to an even greater degree.
These oscillations, as well as those visible in the optimized geometries of Fig.~\ref{potentialfigure} due to the silicon lattice distortions, represent qualitative differences between DFT and effective mass theory.

\begin{figure}
\subfigure[]{\includegraphics[height=3in,width=3in,angle=0.0]{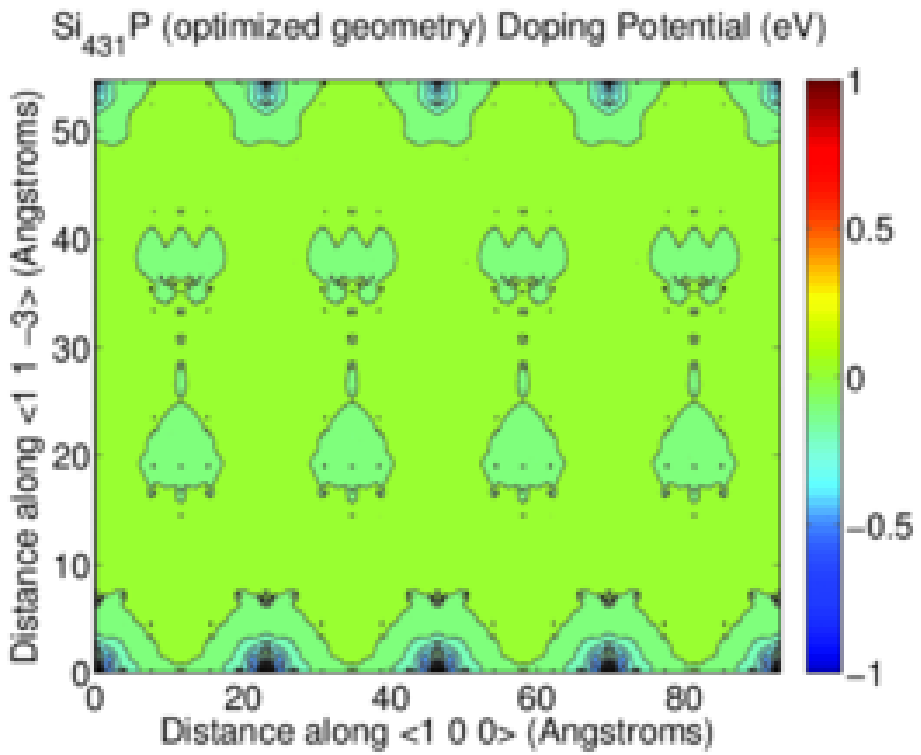}\label{dpsi}}\\
\subfigure[]{\includegraphics[height=3in,width=3in,angle=0.0]{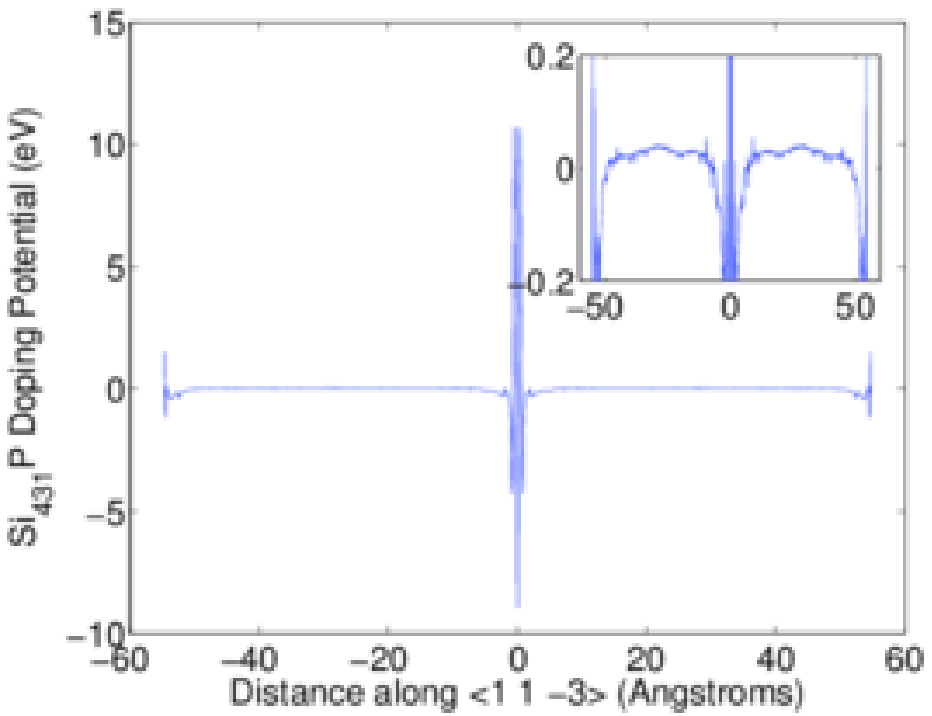}\label{dpmr0}}
\subfigure[]{\includegraphics[height=3in,width=3in,angle=0.0]{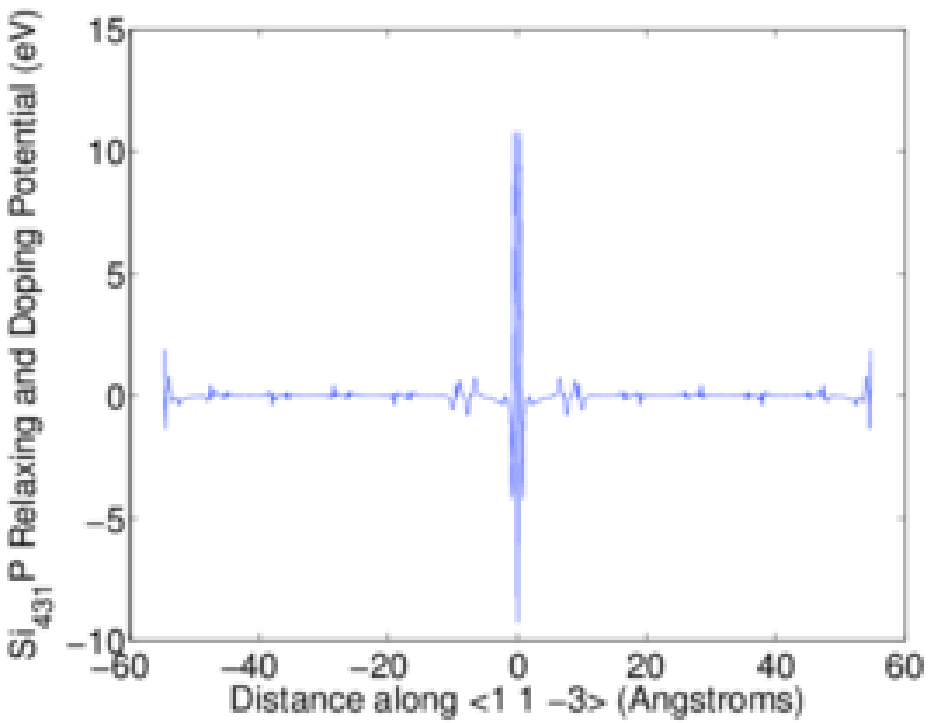}\label{rdpmr0}}
\caption{(Color online) The doping potentials ($V_{Si_{431}P}-V_{Si}$) as a function of distance from the doping layer.  
Fig.~\ref{dpsi} gives a contour plot with the x-direction representing an intralayer direction and the y-direction representing an interlayer direction.  
The contours and color axis are the same as in Fig.~\ref{potentialfigure}.
Fig.~\ref{dpmr0} shows the potential as a function of distance from the doping layer along the line between dopants.
The inset shows an enlargement for the energy range -200 to 200 meV, which is most relevant to scattering by a two-dimensional electron gas.
Fig.~\ref{rdpmr0} shows the potential with geometric relaxation included.\label{dopinglayerfigure}}
\end{figure}

In Ref.~\cite{Carter_Nanotech_11}, Carter \textit{et. al.} use mixed pseudopotentials in order to estimate the potential as a function of the distance from a layer of dopants.
In Fig.~\ref{dopinglayerfigure}, we plot the doping potential (doped minus bulk) as a function of distance from the (001) layer of dopants. 
In Fig.~\ref{dpsi}, a full two-dimensional cut through the potential is given in the plane perpindicular to the (001) plane, and in Figs.~\ref{dpmr0}~and~\ref{rdpmr0} the potential is shown in a slice of this plane which connects two dopants.
In contrast to Ref.~\cite{Carter_Nanotech_11}, where no structure is evident, marked structure is seen in Fig.~\ref{dopinglayerfigure}.
In Figs.~4~and~7 of Ref.~\cite{Carter_Nanotech_11}, the mixed pseudopotential doping potentials are much smoother than in Fig.~\ref{dpmr0}, especially in the region around the dopant.
In Fig.~\ref{dpmr0}, there is a significant amount of structure in the potential near the dopant itself.
Minor effects of the silicon atoms in the next layer of the crystal are also evident in Fig.~\ref{rdpmr0} when the effects of geometric relaxation are included.
It is important to note that these results are for doping densities near the single dopant limit: a study of the effect of a $\delta$-layer of dopants would require very large cells which would likely have thousands of atoms. 
\textcolor{black}{Additionally, the dopant potentials in Ref.~\cite{Carter_Nanotech_11} are plane-averaged, while we have plotted straight point potentials.  
However, averaging does not eliminate the structure in our potentials, but instead reduces the peak potential relative to the somewhat noisy structure of the atomic lattice.}

\subsection{Constrained spin calculations\label{spinsection}}
In the effective mass model~\cite{KohnLuttinger,MacMillen_PRB_84,Kettle_PRB_03,Kettle_JPCM_04,Wellard_JPCM_04,Calderon_PRL_06,Calderon_PRB_07,Hao_PRB_09}, 
the dopants are treated as effective hydrogenic one-electron systems, with spin due to the additional dopant electron.
Density functional theory gives a more detailed treatment of the many-body problem, 
allowing dopant electrons to couple to core electrons on the same dopant, to electrons in the silicon atoms, and 
most importantly, also allowing dopant electrons to couple with each other at low densities.
In order to compare the DFT results with the frequently used single-electron picture~\cite{Morton_Nature_Review,Feher_ESR1,Feher_ESR2,Gordon_PRL_58,Witzel_PRB_06,Abe_PRB_10,Tyryshkin_PRB_03,Schenkel_APL_06}, we calculated the spin densities (spin up density minus spin down density) of each doped cell.

\begin{figure}[htp!]
\subfigure[]{\includegraphics[height=2in,width=2in,angle=0.0]{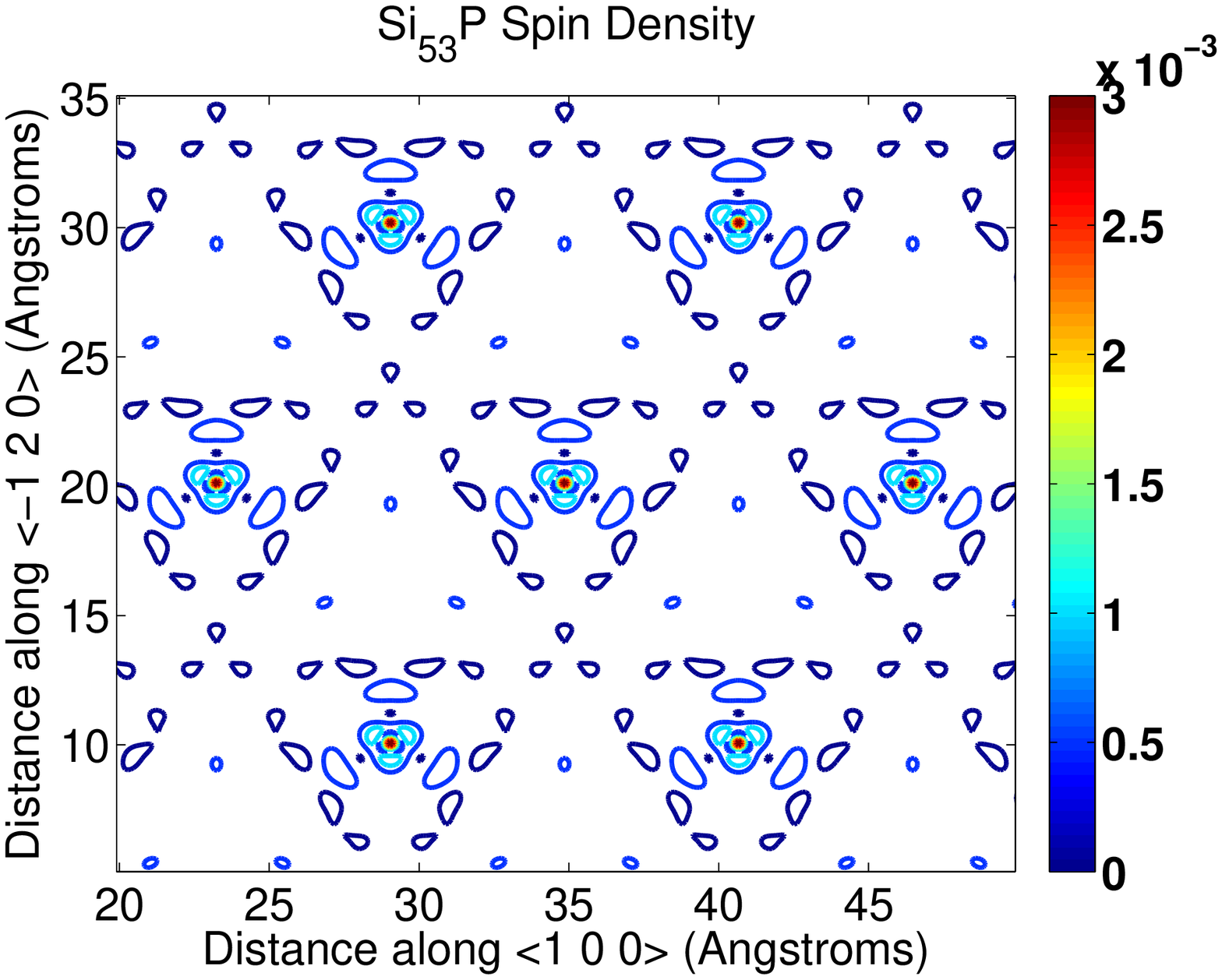}\label{s54}}
\subfigure[]{\includegraphics[height=2in,width=2in,angle=0.0]{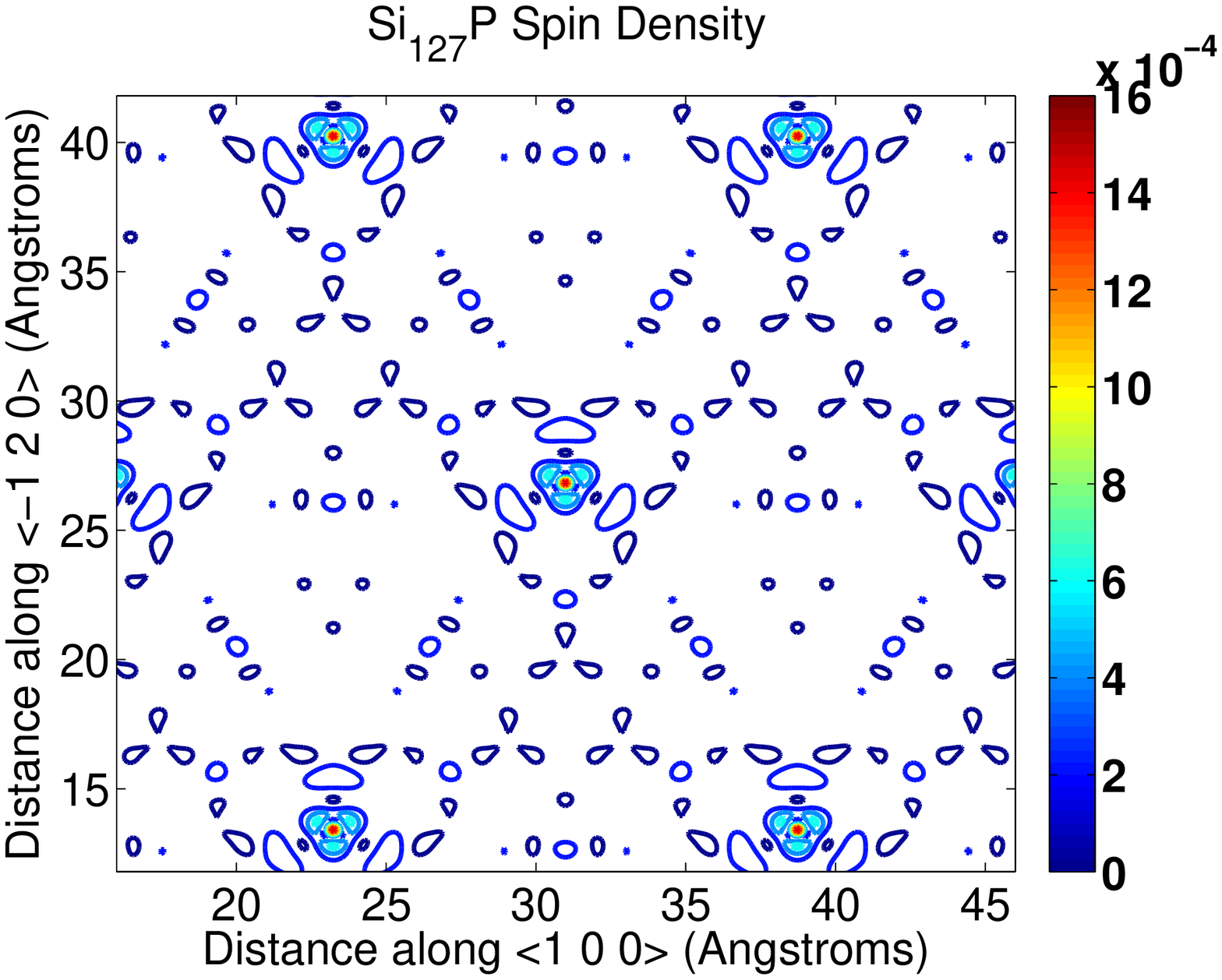}\label{s128}}\\
\subfigure[]{\includegraphics[height=2in,width=2in,angle=0.0]{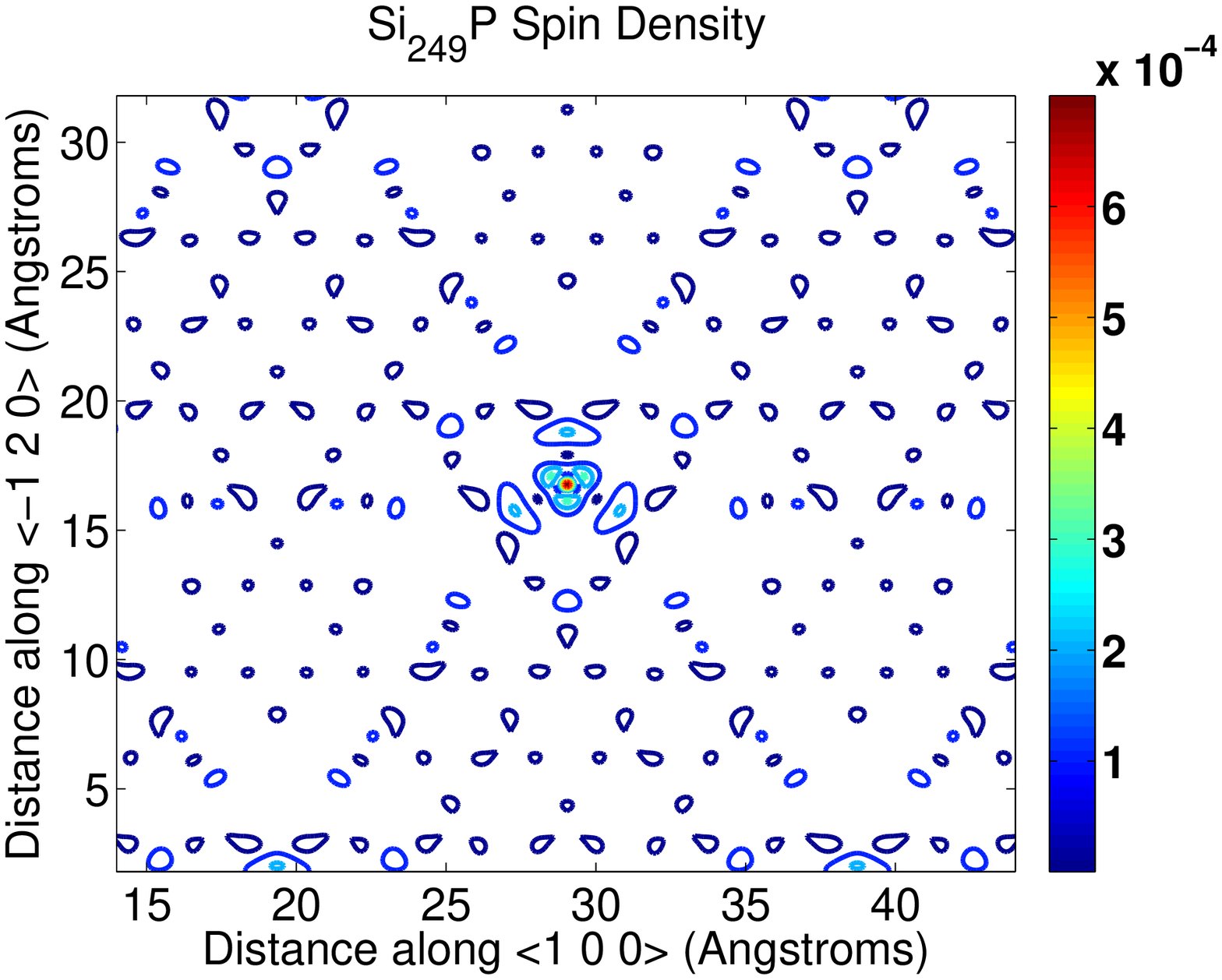}\label{s250}}
\subfigure[]{\includegraphics[height=2in,width=2in,angle=0.0]{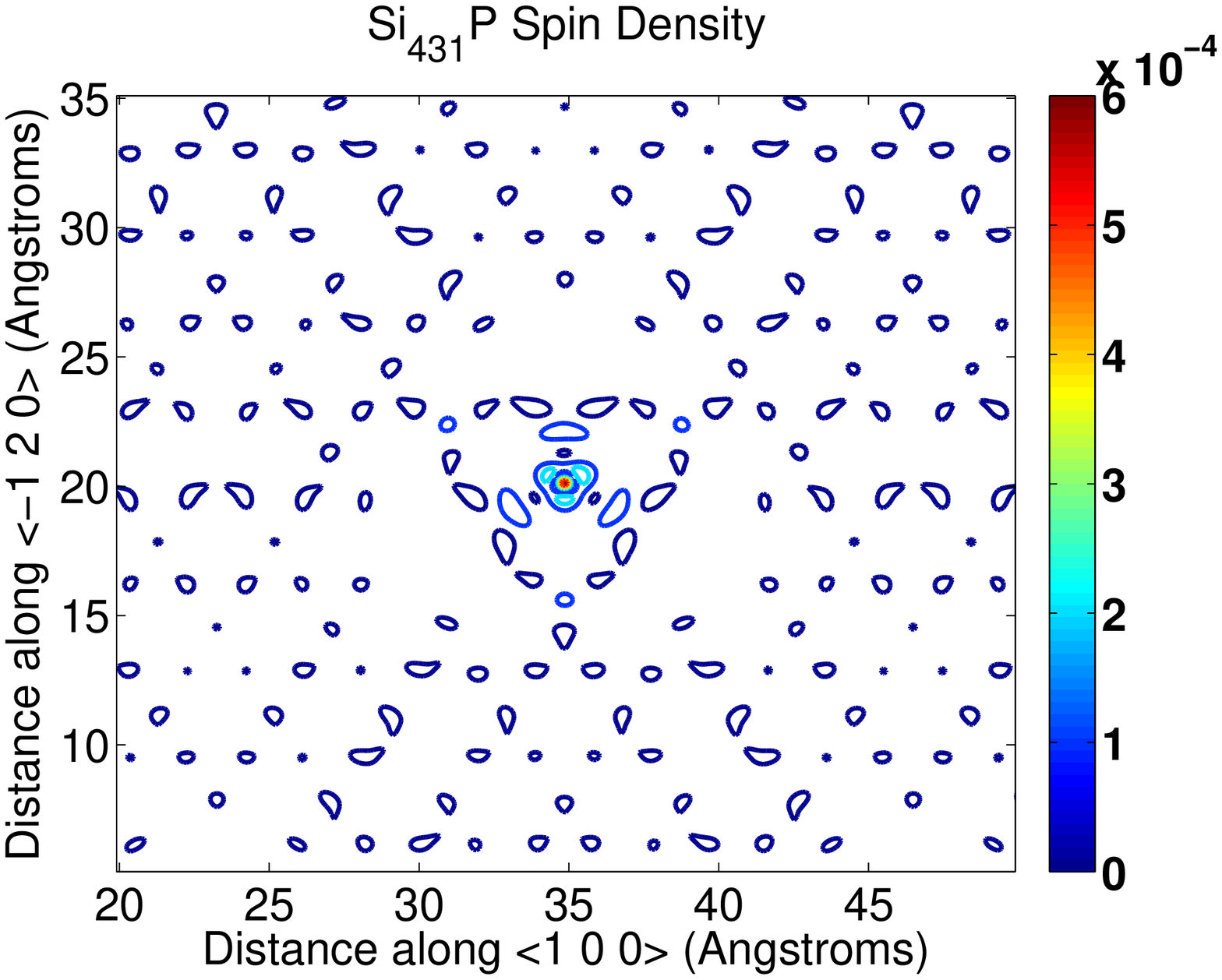}\label{s432}}\\
\subfigure[]{\includegraphics[height=2in,width=2in,angle=0.0]{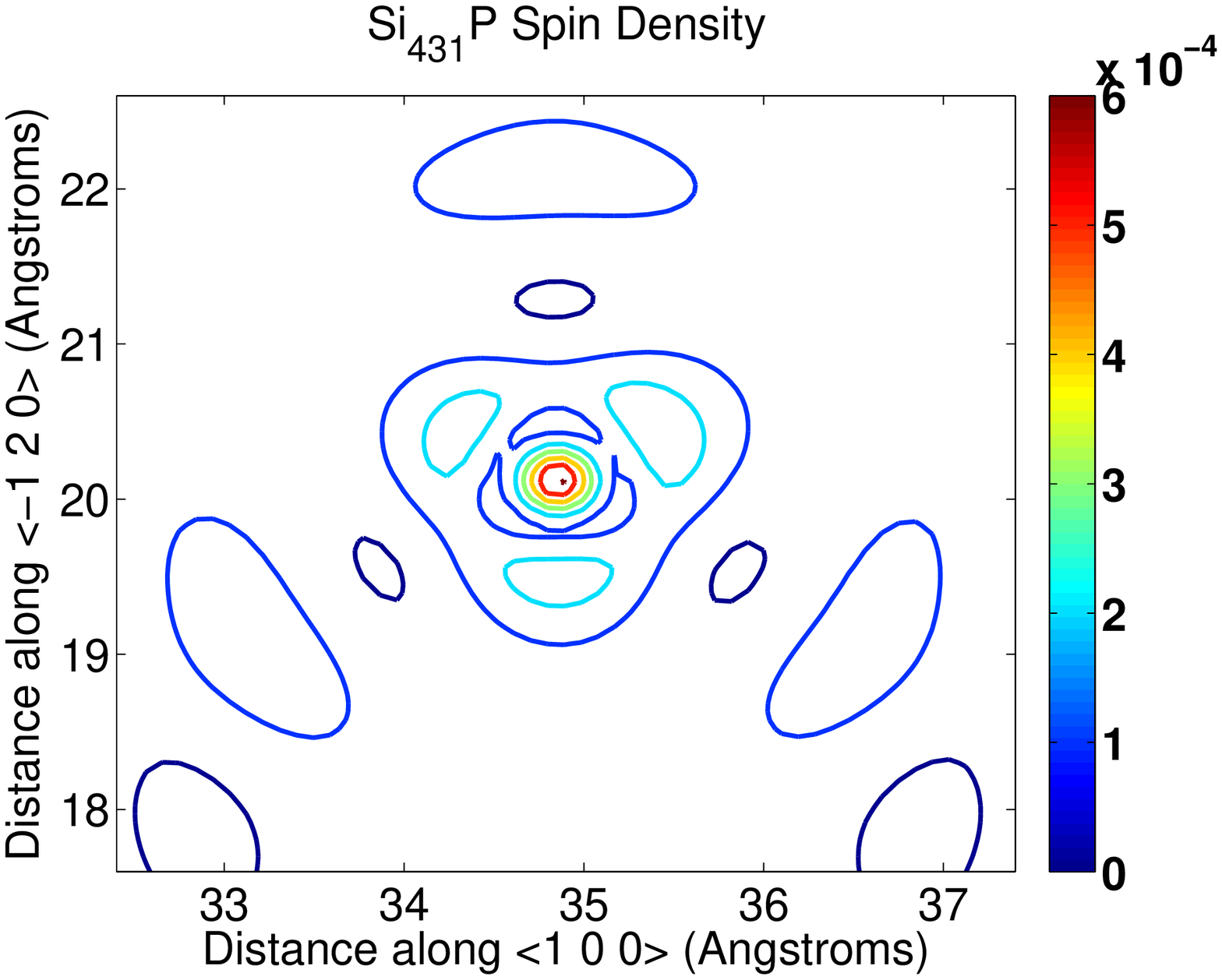}\label{s432cu}}\\
\caption{(Color online) DFT calculations of spin densities ($\rho _\uparrow -\rho _\downarrow$) of the 54, 128, 250, and 432-atom unit cells in a 15 $\times$ 15 \AA~box centered on a dopant, with
the spin constrained to $S=\frac{1}{2}$ \textcolor{black}{along the z-axis} in each unit cell.
A 2.5 $\times$ 2.5 \AA~close up of the dopant area of the spin density for the 432-atom cell is also shown~(\ref{s432cu}).\label{spindensityfigure}}
\end{figure}

Fig.~\ref{spindensityfigure} shows spin densities for DFT calculations in which the total spin of the individual unit cells was constrained to be $S=\frac{1}{2}$ a.u. (one bohr magneton) \textcolor{black}{along the z-axis}.
\textcolor{black}{
The spin density is $\rho _\uparrow -\rho _\downarrow$, where $\rho _\uparrow$ and $\rho _\downarrow$ represent the density of electrons with spin aligned or anti-aligned with the z-axis, respectively.
In all of the previous sections, the calculations were unconstrained in the spin degree of freedom and \textcolor{black}{since the ground state of silicon is not magnetic, these all converged to an average spin of $S=0$, meaning the unpaired electron points in random directions with none preferred over any others.}
While the electron donated by the phosphorus atom does indeed have a spin of $\frac{1}{2}$, in the ground state it can be considered as being in a superposition of its aligned and anti-aligned states, leading to a total spin of $S=0$.
In the following spin-constrained calculations, an effective external field is added in the form of an energy penalty,
\begin{equation}
\label{energypenalty}
E_{total} = E_{DFT}+\lambda \left( \rho _\uparrow - \rho _\downarrow - \frac{1}{2}\hat{z} \right).
\end{equation}
}
In addition to giving 
an indication of spin ordering due to the coupling between dopant atoms in the system with respect to cell size, the local spin density also gives a qualitative picture of where the ``additional'' electron provided by the phosphorus dopant is located
when the total magnetization of the system is locally constrained.  
In the 54 and 128 atom unit cells, the spin density clearly shows interactions between the dopants in the form of areas of high density between dopants.
The 250 and 432 atom cells show comparatively less localization of spin density between the dopant atoms, 
but there are still areas of enhanced spin density connecting the dopants along lines in the (001) plane.  
These areas are fading for the 432 atom cell but not completely absent.
Note also that the maximum spin density is decreasing from the 54 to the 250 atom cell, but remains approximately constant between the 250 and 432 atom cells.
These results give an effective one-electron picture of how modulations in the spin density are affected by the distance between dopant atoms and provide insight on 
the behavior of the effective one-electron wavefunction in this system.

\begin{figure}
\begin{center}
\includegraphics[height=3in,width=3in,angle=0.0]{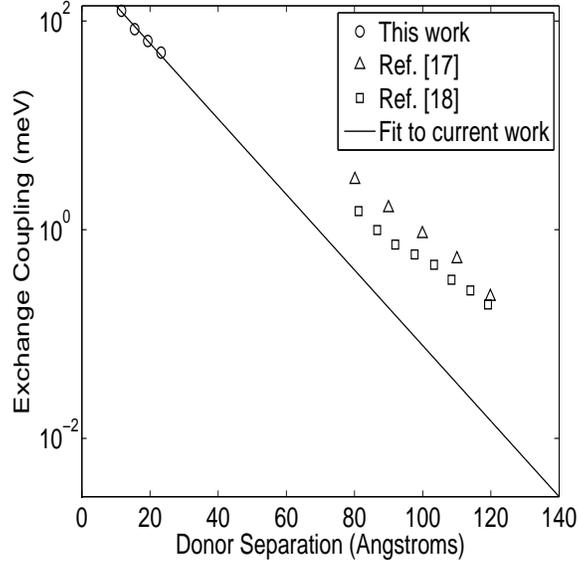}
\end{center}
\caption{Magnitude of the zero-field exchange coupling calculated by DFT in this work (circles), and in the Heitler-London approximation with effective mass theory wavefunctions in Ref.~\cite{Kettle_JPCM_04}~(triangles., Bohr radius of 2.381 nm), and in Ref.~\cite{Wellard_PRB_03}~(squares, Bohr radius of 1.368 nm).  Data for the three largest systems studied here are fit well by a single exponential decay in the spacing between 
donors (solid line.)  The magnitude of the exchange coupling predicted by the fitted decay function for isotropic doping is much lower than that calculated in Refs.~\cite{Kettle_JPCM_04}~or~\cite{Wellard_PRB_03} using the Heitler-London approximation for 
an isolated pair of P atoms in bulk Si (inset.)\label{exchangefigure}}
\end{figure}

An exchange coupling between donors can be estimated using density functional theory according to Eq.~(\ref{jcoupling})\cite{Rudra_JCP_06,Ruiz_JCC_11,Yamaguchi_CPL_79,Noodleman_JACS_85,Noodleman_JCP_81,Illas_TCA_00,Dai_JCP_03}.
The quantity $J$ provides an estimation of the spin couplings between donor qubits which may be used to apply two-qubit gates.
The exchange couplings calculated for the different size supercells are given in Table~\ref{exchangetable}.
\begin{table}[htp!]
\caption{The exchange parameter as a function of cell size.\label{exchangetable}}
\begin{ruledtabular}
\begin{tabular}{cdd}
\multicolumn{1}{c}{Unit Cell}&
\multicolumn{1}{c}{Distance between dopants (\AA)}&
\multicolumn{1}{c}{Exchange coupling (meV)}\\
\hline
Si$_{53}$P&11.62&-125.1\\
Si$_{127}$P&15.49&-83.6\\
Si$_{249}$P&19.36&-64.4\\
Si$_{431}$P&23.23&-49.8
\end{tabular}
\end{ruledtabular}
\end{table}
As the dopant density decreases, we see that the magnitude of the exchange coupling also decreases.   
In Refs.~\cite{Kettle_JPCM_04}~and~\cite{Wellard_JPCM_04}, the exchange coupling of a two-donor system is calculated for dopants at much greater distances than in this work, using a Heitler-London approximation with variable alignment of the longitudinal and transverse Bohr radius of the dopant relative to the inter-dopant direction, respectively.
These works, together with Refs.~\cite{Calderon_PRB_07}~and~\cite{Koiller_PRL_01}, have concluded that oscillations in the exchange coupling would make it difficult to control a quantum information system which attempts to exploit this coupling.  
Although donor spacings studied here are small compared to the previous studies of a pair of phosphorus donor atoms in bulk Si in Refs.~\cite{Kettle_JPCM_04}~and~\cite{Wellard_PRB_03}, we may nonetheless make a comparison of our results with 
these in Fig.~\ref{exchangefigure}.  
For the three largest systems studied here, the behavior of the exchange coupling with respect to donor spacing ($r$) is fit well by a single exponential decay, $J(r)=-323.0\exp(-r/12.0)$.  
The exchange coupling at the distances studied using the 
Heitler-London approximation is systematically larger than the corresponding value extrapolated from our fitted decay.  The source of the decrease in magnitude of the DFT estimated exchange coupling is likely due to oscillations in the donor electron densities  
which are present in the DFT calculation, but which are not included in the models used in 
Refs.~\cite{Kettle_JPCM_04}~and~\cite{Wellard_PRB_03}.  We note that due to the relatively small spacing between donors and the isotropic distribution of donors in this work, the correlations between donor 
atoms are stronger than what is modeled in the studies of isolated pairs of atoms separated by a large distance.  However, if in actual devices the donor atoms are not evenly distributed and well separated, the magnitude of the exchange coupling may also be 
decreased 
with respect to what is predicted by the Heitler-London model.   Extending our studies to larger systems with anisotropic doping will allow for a more direct comparison with studies based of isolated donor pairs.

\subsection{Doping density\label{densitysection}}
\begin{figure}[htp!]
\subfigure[]{\includegraphics[height=3in,width=3in,angle=0.0]{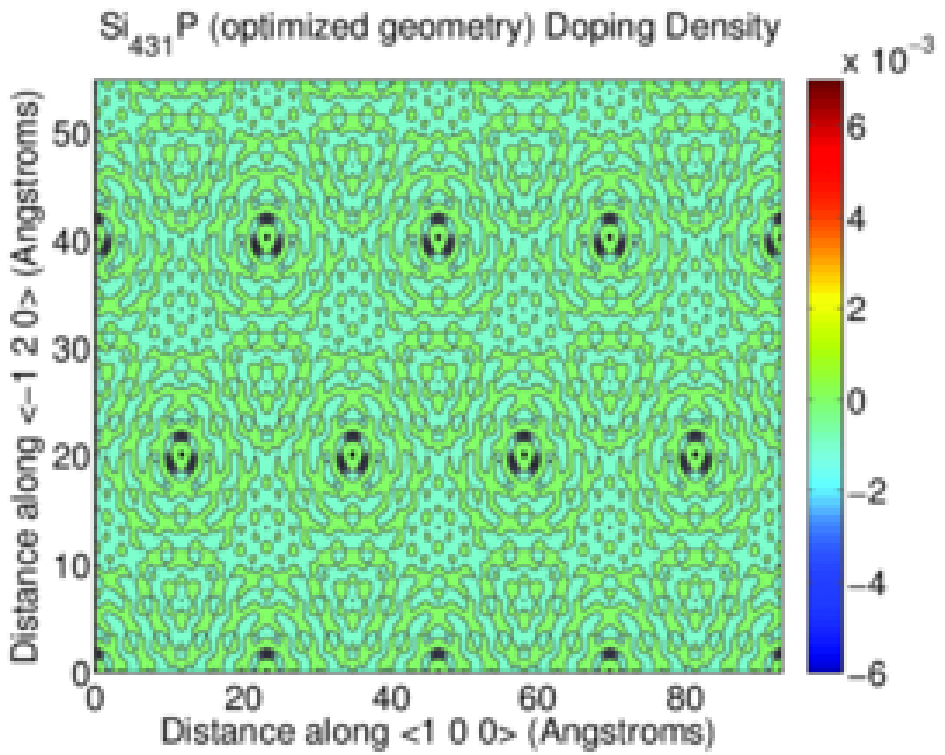}}
\subfigure[]{\includegraphics[height=3in,width=3in,angle=0.0]{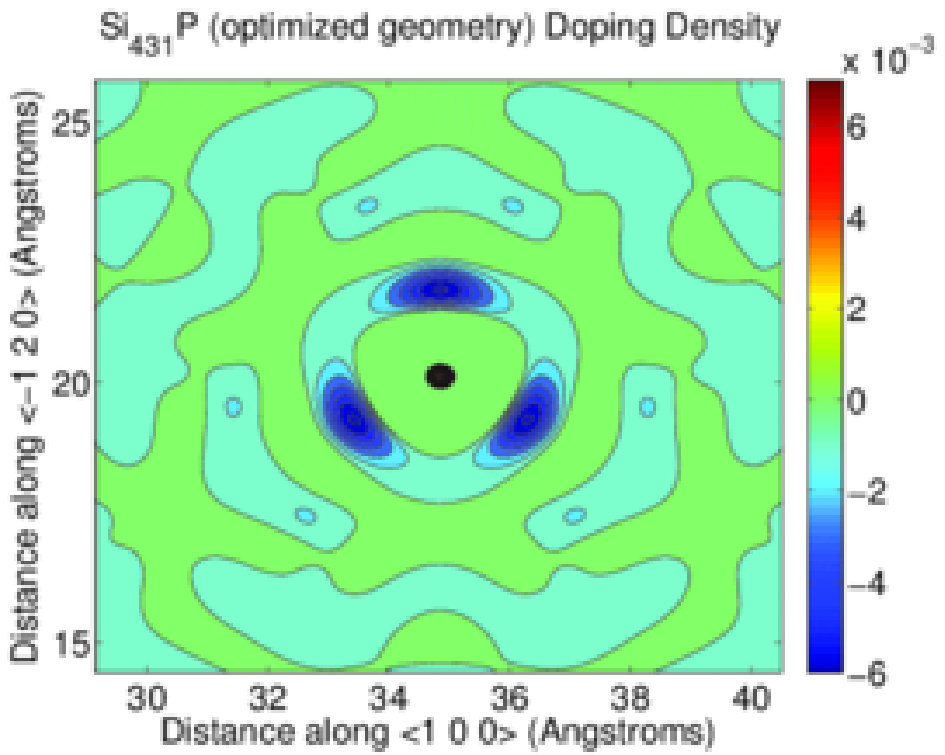}\label{ddzoom}}
\caption{(Color online) The doping density ($\rho _{Si_{431}P}-\rho _{Si}$) in the (001) plane.
The region of the density near the dopant is zoomed in panel~\ref{ddzoom}.
The contours are drawn every 0.01 units, except between -0.1 and 0.0 where they are drawn every 0.002 units.
The color axis is set between -6 and 7 $\times$ 10$^{-3}$ units, while the density varies between -6 $\times$ 10$^{-3}$ and 0.35 units with the larger values near the nucleus.\label{dopantdensityfigure}}
\end{figure}

In Ref.~\cite{Sarma_Heitler_London}, wavefunctions of phosphorus dopants were calculated using effective mass theory, in which the Bloch functions of silicon where taken directly from a density functional theory calculations.
Doping densities (density of doped cell minus undoped cell) were presented in Fig. 2 of Ref~\onlinecite{Sarma_Heitler_London}.
In Fig.~\ref{dopantdensityfigure}, we present the doping density,
\begin{equation}
\rho _{doping} = \rho _{doped}-\rho _{undoped},
\end{equation}
calculated for the 432 atom cell using a full atomistic DFT treatment, where each density $\rho$ is calculated from the Kohn-Sham orbitals $\phi _i$ as
\begin{equation}
\rho = \sum _i \vert \phi _i \vert ^2.
\end{equation}
The current DFT doping densities show a different distribution in the (001) plane than those presented in Ref.~\cite{Sarma_Heitler_London}.
In particular, the present densities are more circular around the dopant in this plane.
The oscillations which are not included in the effective mass calculations are also evident in the immediate vicinity of the dopant.
These oscillations are about two orders of magnitude less than the doping density in the vicinity of the dopants.
Lobes similar to those seen in the doping potentials in Section~\ref{potentialsection} are also apparent in the doping densities.
Finally, we note that the doping density shows evidence of the interaction between dopants, resulting in finite electron density between dopants.

\section{Conclusion}
We have presented density functional theory calculations for 
silicon doped with a single phosphorus atom, for systems with up to 432 atoms in a cell.
A detailed knowledge of the electronic structure of doped silicon is necessary for the implementation of spin-based qubits in silicon~\cite{Kane_Proposal,Skinner_Proposal,Loss_DiVincenzo_Proposal,Levy_JPCM_Review}.
We have calculated non-spherical electron densities and doping potentials which allow for a level of microscopic description beyond effective mass and tight binding theory.
In comparison to previous calculations of dopant electronic structure~\cite{KohnLuttinger,MacMillen_PRB_84,Kettle_PRB_03,Kettle_JPCM_04,Wellard_JPCM_04,Calderon_PRL_06,Calderon_PRB_07,Hao_PRB_09,Fang_PRB_02,Lansbergen_NP_08,Rahman_PRB_09,Rahman_PRB_10,Carter_PRB_09,Carter_PRB_09E,Carter_Nanotech_11}, we have found an unprecedented level of structure in the doping potentials (Fig.~\ref{potentialfigure}) and densities/wavefunctions (Fig.~\ref{dopantdensityfigure}).
\textcolor{black}{Due to the oscillatory nature of doping potentials, the exchange coupling between qubits obtained by extrapolating our results to smaller distances was found to be less than estimates based on the Heitler-London approximation, although further calculations at larger dopant separations are required to confirm this result.}

These calculations have many potential uses.
Such detailed microscopic calculations will allow more accurate and detailed device simulations than are currently possible.
By understanding the effects of modulations in the doping density including effects of both the spin density as well as the doping potential, allows now calculations which probe the readout properties of the qubits, especially using techniques such as EDMR~\cite{Ghosh_PRB_92,Xiao_Nature_04,Lo_PRL_11,vanBeveren_APL_08,Lo_APL_07}.
Additionally, alternative qubits such as excited-state dopants~\cite{Stoneham_JPCM_03} or charged dopant qubits~\cite{Sellier_PRL_06,Calderon_PRB_10,Rahman_PRB_11,Taniguchi_SSC_76} can now be explored with this accurate picture of the electronic structure.
The doping potentials calculated here can provide the starting point for effective one-electron calculations of a dopant electron wavefunction, possibly deformed by some electrostatic gate potential.
They also can guide design of multiple qubit devices by providing effective Hamiltonians or potentials for multiple-qubit geometries.
Finally, the doping potentials provide input for scattering calculations, including calculations in which the current-carrying electrons are confined to a two-dimensional plane to model electrical readout schemes for silicon quantum computation.
The spin densities we have calculated here can also be used to compare with a single-electron picture and to determine the density of the electron which is donated by the phosphorus.

In the future, we plan to look at systems which more accurately represent the experimental devices.
This will require looking at the effect of the silicon dioxide interface and defects at this interface, a considerably more computationally intensive task.
They may also be extended to the calculation of parameters related to the hyperfine interaction.
These DFT calculations may also be coupled with calculations of the spin-dependent scattering~\cite{SCARLET,Lordi_Scattering_PRB_10} of the two-dimensional electron gas as well as quantum control calculations for the implementation of quantum logic operations~\cite{Viola_PRL_99,KBW_Noise_PRA_06,KBW_Noise_PRA_08,KBW_QND_PRB_08,KBW_arXiv_10}.

\begin{acknowledgments}
The authors wish to acknowledge insight due to conversations with Dr. Vincenzo Lordi (Lawrence Livermore National Laboratory).
Additionally, we wish to thank our experimental collaborators in the group of Dr. Thomas Schenkel (Lawrence Berkeley National Laboratory), especially Dr. Cheuk Chi Lo and Christoph Weis, for helpful conversations and direction.  
This work was supported by the UC Lab Fees Research Program under a grant to the University of California, Berkeley and Lawrence Livermore National Laboratory.  It was also performed in part (HDW) under the auspices of the U.S. Department of Energy by Lawrence Livermore National Laboratory under Contract DE-AC52-07NA27344.
\end{acknowledgments}

\bibliographystyle{apsrevM}
\bibliography{SiQC}

\end{document}